\pdfoutput=1
\documentclass[11pt,twoside,a4paper,cmspaper,final,collab]{cms-tdr}
\def\svnVersion{b9b0e25}\def\svnDate{2025/02/10}\def\cmsCernNoTag{CERN-EP-2024-030}\def\cmsCernDate{\today}\def\cmsMessage{Published in the Journal of High Energy Physics as \href{http://dx.doi.org/10.1007/JHEP02(2025)040}{\doi{10.1007/JHEP02(2025)040}.}}
\begin{document}\cmsNoteHeader{B2G-20-004}

\newcommand{\pp}{\ensuremath{\Pp\Pp}\xspace}
\newcommand{\Hbb}{\ensuremath{\PH\to\bbbar}\xspace}
\newcommand{\ttjets}{\ttbar{}+jets\xspace}
\newcommand{\intLumi}{138\fbinv}
\newcommand{\PHH}{\ensuremath{\PH\PH}\xspace}
\newcommand{\Hbbt}{DeepAK8\,tagger\xspace}
\newcommand{\mjj}{\ensuremath{m_{\text{JJ}}}\xspace}
\newcommand{\mjjj}{\ensuremath{m_{\text{Jjj}}}\xspace}
\newcommand{\mjjjs}{\ensuremath{m_{\PH\PH,\text{corr}}}\xspace}
\newcommand{\mH}{\ensuremath{m_{\PH}}\xspace}
\newcommand{\mHH}{\ensuremath{m_{\PH\PH}}\xspace}
\newcommand{\nsub}{\ensuremath{\tau_{21}}\xspace}
\newcommand{\mjjs}{\ensuremath{m_{\PH\PH,\text{corr}}}\xspace}
\newcommand{\mjjred}{\ensuremath{m_{\text{jj,red}}}\xspace}
\newcommand{\mjone}{\ensuremath{m_{\text{J}_{1}}}\xspace}
\newcommand{\mjtwo}{\ensuremath{m_{\text{J}_{2}}}\xspace}
\newcommand{\LambdaR}{\ensuremath{\Lambda_{\text{R}}}\xspace}
\newcommand{\mx}{\ensuremath{m_{\PX}}\xspace}
\newcommand{\Mplbar}{\ensuremath{\overline{M}_\text{Pl}}\xspace}
\newcommand{\TTcat}{HP\xspace}
\newcommand{\LLcat}{LP\xspace}

\cmsNoteHeader{B2G-20-004}

\title{Search for resonant pair production of Higgs bosons in the \texorpdfstring{$\bbbar\bbbar$}{bbbb} final state using large-area jets in proton-proton collisions at \texorpdfstring{$\sqrt{s}=13\TeV$}{sqrt(s)=13 TeV}}

\date{\today}

\abstract{A search is presented for the resonant production of a pair of standard model-like Higgs bosons using data from proton-proton collisions at a centre-of-mass energy of 13\TeV, collected by the CMS experiment at the CERN LHC in 2016--2018, corresponding to an integrated luminosity of 138\fbinv. The final state consists of two \PQb quark-antiquark pairs. The search is conducted in the region of phase space where at least one of the pairs is highly Lorentz-boosted and is reconstructed as a single large-area jet. The other pair may be either similarly merged or resolved, the latter reconstructed using two \PQb-tagged jets. The data are found to be consistent with standard model processes and are interpreted as 95\% confidence level upper limits on the product of the cross sections and the branching fractions of the spin-0 radion and the spin-2 bulk graviton that arise in warped extradimensional models. The limits set are in the range 9.74--0.29\unit{fb} and 4.94--0.19\unit{fb} for a narrow radion and a graviton, respectively, with masses between 1 and 3\TeV. For a radion and for a bulk graviton with widths 10\% of their masses, the limits are in the range 12.5--0.35\unit{fb} and 8.23--0.23\unit{fb}, respectively, for the same masses. These limits result in the exclusion of a narrow-width graviton with a mass below 1.2\TeV, and of narrow and 10\%-width radions with masses below 2.6, and 2.9\TeV, respectively.}

\hypersetup{
pdfauthor={CMS Collaboration},
pdftitle={Search for resonant Higgs boson pair production in the bbbb final state using large-area jets in proton-proton collisions at sqrt(s)=13 TeV},
pdfsubject={CMS},
pdfkeywords={CMS, physics, Higgs, beyond standard model, jets}}

\maketitle

\section{Introduction\label{sec:Introduction}}

In proton-proton ($\Pp\Pp$) collisions at the CERN LHC,
the standard model (SM) production of a pair of Higgs
bosons~\cite{HiggsDiscoveryAtlas,HiggsDiscoveryCMS,CMSHiggsLongPaper} involves
two destructively interfering processes: the production of a virtual Higgs boson 
via a gluon fusion through an internal fermion loop dominated by the top quark,
\PQt, followed by an $\PH\PH\PH$ vertex, and a `box' Feynman diagram with a 
fermion loop resulting in two $\PQt\PQt\PH$ vertices. 
Its predicted cross section of $30.8^{+6.4}_{-23.1}$\unit{fb} at a centre-of-mass
energy of 13\TeV~\cite{Dawson:1998py,
Borowka:2016ehy,
Baglio:2018lrj,
Shao:2013bz,
deFlorian:2015moa,
Grazzini:2018bsd,
Baglio:2020wgt,
deFlorian:2016spz,
deFlorian:2013jea,
Baglio:2012np} is
too small to be observable with the current data. However, according to many
models ``beyond the SM'' (BSM), other modes of Higgs boson pair production
could exist, many involving the production of a massive BSM resonance \PX that
then decays to a Higgs boson pair ($\PX\to\PH\PH$).

Models with a warped extra dimension (WED), as proposed by Randall and
Sundrum~\cite{Randall:1999ee,RS-1999}, are among the BSM scenarios that
predict the existence of resonances with large couplings to the SM Higgs boson,
such as a spin-0 radion~\cite{Goldberger:1999uk,DeWolfe:1999cp,Csaki:1999mp}
and a spin-2 first Kaluza--Klein (KK) excitation of the
graviton~\cite{Davoudiasl:1999jd,Csaki:2000zn, Agashe:2007zd}. The WED
models~\cite{Giudice:2000av} postulate an additional spatial dimension $l$
compactified between two four-dimensional hypersurfaces known as branes,
with the region in between, the bulk, warped by an exponential metric $\kappa l$,
where $\kappa$ is the warp factor. A value of $\kappa l\!\!\approx\!\!35$ reproduces
the mass hierarchy between the Planck scale \Mpl and the electroweak
scale~\cite{Randall:1999ee}. One of the parameters of the model is
$\kappa/\Mplbar$, where $\Mplbar \equiv \Mpl /\sqrt{8\pi}$. The
ultraviolet cutoff scale of the model $\LambdaR \equiv \sqrt{6} \re^{-\kappa l}
	\Mplbar$~\cite{Goldberger:1999uk} is another parameter, and its value is expected
to be near the TeV scale.

Searches for $\PH\PH$ production have been performed by the ATLAS
\cite{ATLAS:2023vdy,
Aad:2014yja,Aad:2015uka,Aad:2015xja,Aaboud:2016xco,
ATLAS_HH4b_2016_13TeV,ATLAS_bbtautau_2016_13TeV,ATLAS_HHbbgamgam_2016_13TeV,
ATLAS:2021ifb,ATLAS:2022xzm,ATLAS:2022hwc,ATLAS:2024lsk}
and CMS
\cite{CMS:2024phk,
Khachatryan:2014jya,Sirunyan:2017tqo,Khachatryan:2015year,
Khachatryan:2015tha,Khachatryan:2016sey,
Khachatryan:2016cfa,CMSHHbbtatau13TeV2016data,
Sirunyan:2017guj,CMS-HIG-17-008,CMS-HIG-17-009,
CMS:2017aza,CMS:2018kaz,CMS:2018vjd,CMS:2021roc,CMS:2023boe,CMS:2024rgy}
Collaborations using LHC $\Pp\Pp$ collision data at $\sqrt{s}=8$ and
13\TeV.

A search for a KK bulk graviton or a radion decaying to $\PH\PH$ in the
$\bbbar\bbbar$ final state was performed by CMS~\cite{CMS-HIG-17-009} using
events with four separate \PQb quark jets. A similar search targeting a
higher \mx range, in which two large-area jets were used to reconstruct the
highly Lorentz-boosted Higgs bosons has also been published by the CMS
Collaboration~\cite{CMS:2017aza}.
The configuration of a Higgs boson candidate reconstructed as
one large-area jet or as two separate narrow jets depends on its momentum~\cite{Gouzevitch:2013qca}.
A search for a new resonance $X$ decaying to a Higgs boson and a scalar  
$Y$ in the $\bbbar\bbbar$ final state~\cite{CMS:2022suh} targeted 
the same dijet topology in a similar kinematic regime 
but was optimised for a wider range of $Y$ masses, and probed the 
next-to-minimal supersymmetric SM and an extension of the SM 
with two additional singlet scalar fields.

In this paper, we improve upon the CMS search for a high-mass resonance
($1 \le \mx \le 3\TeV$) decaying to
$\PH\PH \to \bbbar\bbbar$~\cite{CMS:2018vjd} by using
data collected at $\sqrt{s} = 13\TeV$ in 2016--2018, 
corresponding to the full Run~2 integrated luminosity of \intLumi.
We use the scenario of
Ref.~\cite{Fitzpatrick:2007qr} to describe the KK graviton, where the
propagation of SM fields is allowed in the bulk and follows the
characteristics of the SM gauge group, with the right-handed top quark
localised near the \TeV brane.
The theoretical values of $\sigma(\Pp\Pp\to\PX)\mathcal{B}(\PX\to \PH\PH\to
	\bbbar\bbbar)$ are calculated for various masses, using
$\LambdaR=3\TeV$ for the radions and
$\kappa/\Mplbar=0.5$ for the bulk gravitons.  For
these values of $\kappa/\Mplbar$ and \LambdaR, the branching fractions
$\mathcal{B}(\PX\to \PH\PH\to \bbbar\bbbar)$ are 10\% and 23\%, for the
bulk graviton and the radion, respectively, for masses of 1~\TeV and larger (cf.
Figs.~4.5 and 4.10 from Ref.~\cite{Oliveira:2014kla}).

Owing to the broad mass range explored, the \Hbb decay is studied using
two analysis topologies. If \mx is large, both Higgs bosons are highly
Lorentz-boosted and are reconstructed using large-area jets. These ``fully-merged''
events are then divided into two categories according to their purity. To identify the
merged \Hbb decays, referred to henceforth as ``\PH candidate jets'', we use a deep neural network jet classifier (``tagger'')
algorithm, described in Section~\ref{sec:EvtSel}.
For resonances with masses in the intermediate range ($0.8$--$1.5$\TeV), the
less energetic Higgs boson often does not produce a merged $\bbbar$ jet, and
thus these events are reconstructed using one large-area jet and a
combination of two separate \PQb quark jets (``semi-resolved'' category).
The inclusion of the semi-resolved events leads to an improvement in
the search sensitivity for resonances with $\mx$ around $1$\TeV.

The two dominant sources of the SM background are multijet
production and top quark pair production in association with jets,
referred to here as \ttjets.  Both backgrounds are estimated from data, but
the procedures are assisted by simulations. To predict the multijet
background, the events that fail the \Hbb identification of the leading-\pt jet
are also used. To aid in the modelling of the \ttjets background, two
categories enriched in \ttjets are defined in addition to three signal categories.
For all five categories, each composed
of two regions with events that pass and fail the \Hbb jet identification, the
background estimation (described in Section~\ref{sec:SigModelBkgEst}) is based on
a two-dimensional fit of the reconstructed resonance mass and
the mass of the leading-\pt large-area jet. In this joint binned likelihood fit
of ten regions the signal strength floats unconstrained, and the nuisance
parameters governing the corrections to both multijet and \ttjets backgrounds are
floating within allowed ranges.  Thus, the signal extraction and the
entire background estimation are done simultaneously.

This paper is organised as follows: a brief description of the CMS detector is
given in Section~\ref{sec:CMSDetector} followed by a description of event
simulation in Section~\ref{sec:Simulation}. The event selection criteria are
defined in Section~\ref{sec:EvtSel}, and
Section~\ref{sec:SigModelBkgEst} describes the modelling of the major
background processes. These are followed by Section~\ref{sec:Systematics} on
the relevant sources of systematic uncertainty and their variations allowed
by the fit.  Finally, the results are presented in Section~\ref{sec:Results}.

Tabulated results are provided in the HEPData record for this analysis~\cite{hepdata}.

\section{The CMS detector and event reconstruction\label{sec:CMSDetector}}

The central feature of the CMS apparatus is a superconducting solenoid of
6\unit{m} internal diameter, providing a magnetic field of 3.8\unit{T}. Within
the solenoid volume are a silicon pixel and strip tracker, a lead tungstate
crystal electromagnetic calorimeter (ECAL), and a brass and scintillator hadron
calorimeter (HCAL), each composed of a barrel and two endcap sections. Forward
calorimeters, made of steel and quartz fibres, extend the pseudorapidity
($\eta$) coverage provided by the barrel and endcap detectors. Muons are
detected in gas-ionization chambers embedded in the steel flux-return yoke
outside the solenoid. A more detailed description of the CMS detector, together
with a definition of the coordinate system used and the relevant kinematic
variables, can be found in Ref.~\cite{CMSdetector}.

Events of interest are selected using a two-tiered trigger system. The first level (L1),
composed of custom hardware processors, uses information from the calorimeters and muon
detectors to select events at a rate of around 100\unit{kHz} within a fixed latency of
4\mus~\cite{CMS:2020cmk}. The second level, known as the high-level trigger (HLT),
consists of a farm of processors running a version of the full event reconstruction
software optimised for fast processing, and reduces the event rate to
around 1\unit{kHz} before data storage~\cite{CMS:2016ngn}.

A particle flow (PF) algorithm~\cite{ParticleFlow} aims to reconstruct
and identify each individual particle in an event (PF candidate),
with an optimised combination of information from the various elements of the CMS detector.
The energy of photons is obtained from the ECAL measurement.
The energy of electrons is determined from a combination of the track momentum at the primary interaction vertex,
the corresponding ECAL cluster energy, and the energy sum of all bremsstrahlung photons attached to the track.
The momentum of muons is obtained from the curvature of the corresponding track.
The energy of charged hadrons is determined from a combination of their momentum
measured in the tracker and the matching ECAL and HCAL energy deposits,
corrected for zero-suppression effects
and for the response function of the calorimeters to hadronic showers.
Finally, the energy of neutral hadrons is obtained
from the corresponding corrected ECAL and HCAL energies.
The primary vertex (PV) is taken to be the vertex corresponding to the hardest scattering in the event,
evaluated using tracking information alone, as described in Section 9.4.1 of Ref.~\cite{CMS-TDR-15-02}.

For each event, jets are clustered from these reconstructed particles
using the anti-\kt
algorithm~\cite{Cacciari:2008gp, Cacciari:2011ma} with a distance
parameter of 0.4 (AK4 jets) or 0.8 (AK8 jets). Jet momentum is
determined as the vectorial sum of all particle momenta in the jet,
and is found from simulation to be, on average, within 5 to 10\% of
the true momentum over the whole transverse momentum (\pt) spectrum
and detector acceptance. Additional $\Pp\Pp$ interactions within the
same or nearby bunch crossings (pileup) can contribute additional tracks and
calorimetric energy depositions, increasing the apparent jet
momentum. To mitigate this effect, tracks identified to be originating
from pileup vertices are discarded, and an offset correction is applied
to correct for remaining contributions~\cite{Sirunyan:2020foa,Bertolini:2014bba}.
Jet energy corrections are
derived from simulation studies so that the average measured energy of
jets becomes identical to that of particle level jets. In situ
measurements of the momentum balance in dijet, photon+jet, $\PZ$+jet,
and multijet events are used to determine any residual differences
between the jet energy scale in data and in simulation, and
appropriate corrections are
made~\cite{Khachatryan:2016kdb}. Additional selection criteria are
applied to each jet to remove jets arising from
instrumental effects or reconstruction failures~\cite{CMS-PAS-JME-16-003}. The jet energy
resolution amounts typically to 15--20\% at 30\GeV, 10\% at 100\GeV,
and 5\% at 1\TeV~\cite{Khachatryan:2016kdb}.

\section{Event simulation\label{sec:Simulation}}

Two scenarios of bulk graviton and radion signal events are considered:
``narrow-width'' signal shapes with a width of 1\MeV and a ``10\%-width'' 
where the width is set to 10\% of the resonance mass.
The width of 1\MeV is much smaller than the standard deviation of the dijet
invariant mass distribution ($\sim$5\%, in the considered mass range 1--3
\TeV) and thus the narrow-width scenario provides a limiting case where the signal
shape is completely determined by the experimental resolution.
A width of 10\% was chosen to demonstrate the effects of a resonance width
larger than the jet mass resolution. Although the background estimation would be
valid for much broader signals, the analysis is not optimised for them.
All signals are simulated at leading order (LO) in
the mass range 1--3\TeV, using the {\MGvATNLO} event generator~\cite{MG5_aMCNLO}; version 2.2.2
is used for the 2016 data-taking period, and 2.4.2 for 2017 and 2018. The
	{NNPDF3.0} LO parton distribution function (PDF)
set~\cite{Ball:2014uwa}, taken from {LHAPDF6} library
~\cite{Harland-Lang:2014zoa,Buckley:2014ana,Carrazza:2015hva,Butterworth:2015oua},
with the four-flavour scheme, is used for all signal samples.
The parton shower and hadronization
are simulated with \PYTHIA~8.212~\cite{Pythia8p2}.

The dominant background consists of events composed primarily of jets (multijet
events) arising from the SM quantum chromodynamics (QCD) interaction, and is
modelled from data. The \ttjets events comprise most of the remaining
background and are generated at next-to-LO using
	{\POWHEG~2.0}~\cite{POWHEG, POWHEG_Frixione, POWHEGBOX} using {NNPDF3.0} LO
PDF set for 2016 data-taking period, and {PDF4LHC15} next-to-next-to-LO (NNLO) PDF
set~\cite{Butterworth:2015oua, Dulat:2015mca, Harland-Lang:2014zoa, Ball:2014uwa, Gao:2013bia, Carrazza:2015aoa}
to model data from 2017 and 2018.
These events are showered by
\PYTHIA~8, using the {\sc CUETP8M2T4} tune~\cite{Skands:2014pea,CMS-PAS-TOP-17-007}.
The contribution from \ttjets is estimated using a NNLO cross section of
$832^{+46}_{-52}\unit{pb}$~\cite{Czakon:2011xx}, corresponding to the top quark
mass of 172.5\GeV.
To account for the difference in the shape of the \pt distribution of the top
quarks between data and simulation, arising from the absence of the contribution from 
the NNLO diagrams, the simulated \ttjets events are reweighted using the \pt-dependent 
scale factor, $\re^{\alpha - \beta \pt}$, with $\alpha = 0.0615$ and $\beta = 0.0005$, derived 
from low-\pt \ttjets events.
A sample of multijet events from QCD interactions, simulated
at LO using \MGvATNLO and \PYTHIA~8, and {NNPDF3.0} (for 2016)
or {PDF4LHC15} (for 2017 and 2018) is used to develop and validate
the background estimation techniques prior to being applied to the data. Other
background processes, such as $\PW\PZ$, $\ttbar\PZ$ or $\PZ$+jets production, are also considered
but their yields are found to be negligible.

All generated samples are processed through a
\GEANTfour-based~\cite{Agostinelli:2002hh,GEANT} simulation of the CMS detector.
The effect of pileup, averaging 23--32 additional interactions per bunch crossing,
for the LHC beam conditions between
2016 and 2018, is included in the simulations, and the samples are reweighted to
match the distribution of the number of $\Pp\Pp$ interactions observed in the
data, assuming a total inelastic $\Pp\Pp$ collision cross section of
69.2\unit{mb}~\cite{CMSLumi13TeV}.

\section{Event selection\label{sec:EvtSel}}

Collision events are selected using a logical OR of triggers based on the jet
activity in the event. One trigger path requires that the \pt sum of all AK4
jets in the event (\HT) be greater than 800, 900, or 1050\GeV, depending on the
data collection year and the LHC beam instantaneous luminosity.
A second trigger path collects events with
$\HT > 650\GeV$, and with a pair of AK4 jets that has invariant mass above 900\GeV and a
pseudorapidity separation $\abs{\Delta\eta}<1.5$. A third
trigger path accepts events if the \pt of the leading AK8 jet is greater than 360
or 400\GeV (depending on the data collection year) and the ``trimmed mass'' of an AK8 jet is above 30\GeV. The jet
trimmed mass is obtained after removing remnants of soft radiation with the jet
trimming technique~\cite{Krohn:2009th}, using a subjet size parameter of 0.3
and a subjet-to-AK8 jet \pt fraction of 0.1.

Offline, collected events are
split into three categories: one semi-resolved category and two fully-merged
categories, further separated by purity.
Since the background estimation uses the mass and the \Hbb jet tagger discriminant
of the leading-\pt AK8 jet, the events are not preselected based on these variables.
The AK8 jets are required to have $\abs{\eta}<2.4$, and
$\pt > 300\GeV$. The fully-merged categories require two such AK8
jets (each representing a Higgs boson candidate), whereas the semi-resolved category
requires only one, with the
other Higgs boson candidate reconstructed from a pair of \PQb-tagged AK4 jets.

A resonant $\PH\PH$ signal of high mass results in a small
$\abs{\Delta\eta}$ between the two Higgs bosons, while the
multijet background often produces events with larger
values of $\abs{\Delta\eta}$. Events in the fully-merged category are
therefore required to have $\abs{\Delta\eta} < 1.3$ between the \PH candidate jets.
The subleading AK8 jet is required to have its soft-drop mass, the jet mass that results from applying the soft-drop
algorithm~\cite{Dasgupta:2013ihk,Larkoski:2014wba}, between 110--140\GeV,
consistent with the Higgs boson mass, $\mH=125\GeV$~\cite{Aad:2015zhl,Sirunyan:2017exp}.
The fully-merged selection is summarised in Table~\ref{tab:table1plus1}.

\begin{table}[ht]
	\centering
		\topcaption{Event selection criteria for the fully-merged topology.}
		\label{tab:table1plus1}
		\begin{tabular}{ l c }
			\hline
			Variable                           & Selection                               \\
			\hline
			Leading two AK8 jets               & $\pt > 300 \GeV$ and $\abs{\eta} < 2.4$ \\
			$\abs{\Delta\eta}$                 & $<$1.3                                 \\
			Sub-leading AK8 jet soft-drop mass & $110 < m_\text{SD} < 140 \GeV$          \\
			$\mjjs$                            & $>$750\GeV                            \\
			\hline
		\end{tabular}
\end{table}

A deep neural network based tagger, ``DeepAK8''~\cite{DeepAK8}, is used to
identify the boosted \Hbb candidate jets. We use a ``mass-decorrelated''
version of this tagger, which exploits an adversarial network to reduce 
the correlation of the tagging score with the soft-drop jet mass~\cite{DeepAK8}.
A significantly reduced sculpting of the distribution of the \PH candidate's 
jet mass preserves its sidebands and allows the use of the jet mass 
in conjunction with the \Hbbt in the background estimate.
The efficiency of the \Hbbt is calibrated in data using a sample of
jets originating from gluons splitting into $\bbbar$ pairs that produce merged
jets. A jet in this sample must have a soft-drop mass in the 50-200 \GeV range and
have two secondary vertices, each matched to one subjet. This selection results
in tagger distributions that are similar between the sample jets and the signal
\Hbb jets. The \Hbbt data-to-simulation correction factors range from
0.9 to 1.4, depending on the jet \pt and data-taking year~\cite{Sirunyan:2017ezt}.
The \Hbbt outperforms the ``double-b'' \Hbb tagger used previously~\cite{CMS:2018vjd},
resulting in an increase of the sensitivity from the tagger alone 
by a factor of $\approx$2.5 over the whole search domain.

The AK8 jets with \Hbbt discriminant above $0.8$ are said to pass a ``loose''
criterion while those with the discriminant above $0.9$ pass the ``tight''
criterion. The efficiency of the tight criterion for \PH candidate jets from a
1500\GeV narrow radion signal is about 60\%, with a misidentification
probability of QCD jets of 1\%.  For jets that pass the loose but not tight
criterion, the \PH jet efficiency is about 20\%, with the misidentification
probability of 2\%.
The fully-merged events are split into two categories based on
the purity of the \PH candidate jets: events are
categorised as either ``high purity'' (\TTcat), where both AK8 jets satisfy the tight
threshold, or as ``low purity'' (\LLcat), where both AK8 jets pass the loose 
tagging threshold but are not part of the \TTcat category.

\begin{figure}[tbh!]
	\centering
	\includegraphics[width=0.6\textwidth]{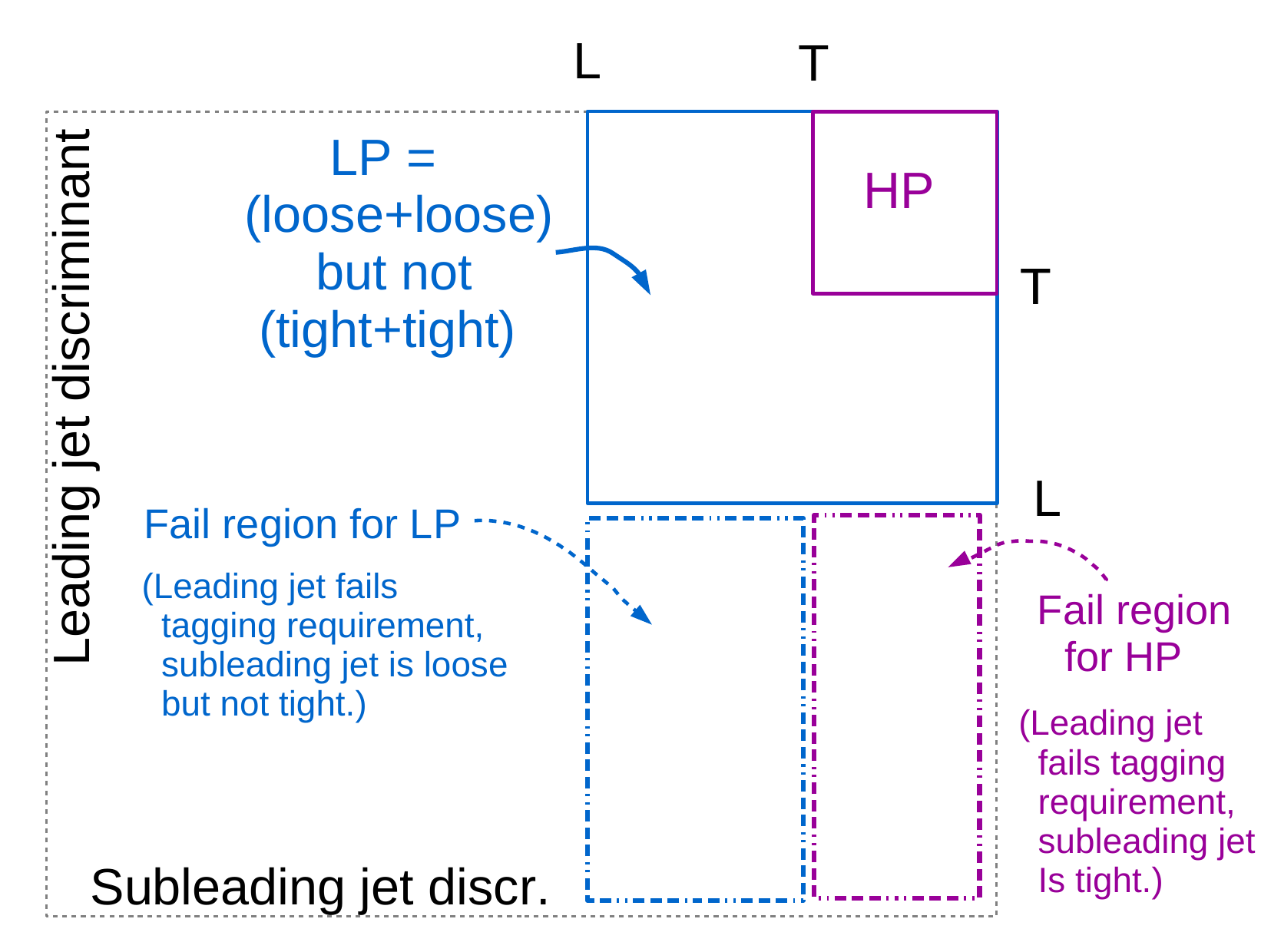}
	\caption{A diagram showing high-purity (HP, purple) and low-purity (LP, blue) pass regions 
	    (solid) and their corresponding fail regions (dash-dotted).}
	\label{fig:HH_AT_considerations}
\end{figure}

We denote the signal regions as ``pass'' regions.
For the purpose of background estimation, for each signal region
we also define a control region where the leading-\pt AK8 jet fails the tagging
requirement; we denote them as ``fail'' regions, and define them separately for
\TTcat and \LLcat categories.
In defining the mutually exclusive \TTcat and \LLcat fail regions, we aim to
model the signal regions with events that have the same
criteria for the subleading jet, which makes them kinematically similar.
The \TTcat fail region (used to predict the background in the \TTcat signal region) is defined by the
leading-\pt \PH candidate jet failing the loose tagger requirement, while the subleading-\pt \PH candidate
jet passes the tight \Hbbt requirement.
Analogously, the \LLcat fail region is defined by
the leading \PH candidate jet failing the loose criterion
while the subleading passes it, but fails the tight one.
A schematic diagram of these four regions is shown in Fig.~\ref{fig:HH_AT_considerations}.
The \TTcat selection corresponds to a signal efficiency of 7--11\% for a
narrow radion signal for masses \mx in the range 1--3\TeV, and
slightly higher for the bulk graviton.
The \LLcat selection results in signal efficiencies of 3--4\% over
the same \mx domain.

Two dedicated \ttjets event control regions (each consisting of corresponding 
\TTcat and \LLcat events) are also used to correct the modelling of the \ttjets
background component for events with high jet \pt, for which the \ttbar
simulation does not agree with data. The \ttbar
control regions use the same selections as the \TTcat and \LLcat categories,
except for a window on the soft-drop mass of the subleading-\pt jet, which is
shifted from $110 < m_\text{SD} < 140$ to $140 < m_\text{SD} < 210\GeV$ in order to
correspond to the top quark mass.

Events that fail the fully-merged selection for either \TTcat or \LLcat category
are considered in the semi-resolved selection.
Jets for the semi-resolved category are required to have $\abs{\eta}<2.4$,
and $\pt > 30\GeV$ ($300\GeV$) for AK4 (AK8) jets.

To find a Higgs boson decay into two resolved \PQb quark jets, all
AK4 jets in each event are examined by the
``DeepJet'' algorithm~\cite{Bols:2020bkb,CMS-DP-2018-058}, which gives
the probability for a jet to have originated from a bottom quark.
DeepJet is a neural network trained using
information from tracks and secondary vertices associated with the jet.

The DeepJet selection on AK4 jets uses the ``medium'' working point,
which corresponds to a 1\% mistag rate for gluon and light-flavoured quark jets.
It results in
a \PQb tagging efficiency of about 70\% for \PQb quark jets in
the \pt range 80--150\GeV, and decreasing to about 50\% for
$\pt\!\approx\!1000\GeV$. The \PQb tagging efficiency in the
simulation is corrected to match that in the data, using
measurements of the \PQb-tagging algorithm performance in a sample
of muon-enriched jets and \PQb jets from \ttjets events, with the
correction factor ranging from 0.95 to 1.1\cite{CMS-DP-2018-058}.

Resolved \Hbb candidates are constructed by considering all pairs of
\PQb-tagged AK4 jets. Events are required to have least one
pair where both AK4 jets (jets ``$\text{j}_2$'' and ``$\text{j}_3$'')
are separated by
$\Delta R = \sqrt{\smash[b]{(\Delta\phi)^2 + (\Delta\eta)^2}}  > 0.8$
(where $\phi$ is the azimuthal angle in radians) from the
leading-\pt AK8 jet (jet ``$\text{J}_1$'') and are within $\Delta R < 1.5$ of each other.
If several such pairs are found, the pair of jets $\text{j}_{2}$ and
$\text{j}_{3}$ that has the highest sum of the AK4 jet DeepJet
discriminant values is selected. The invariant mass of $\text{j}_{2}$ and
$\text{j}_{3}$, $m_{\text{j}_{2} \text{j}_{3}}$, is required to be
within 90--140\GeV, forming the resolved \Hbb candidate.
The leading-\pt AK8 jet is then
identified as the merged \PH candidate, and the pair of AK4 jets is
identified as the resolved \PH candidate. If no resolved \PH candidate
is found starting from the leading-\pt AK8 jet, then this
process is repeated with the subleading-\pt AK8 jet as a merged \PH candidate.
The event is rejected if a \Hbb pair is not found even in this case.
As in the fully-merged regime, the events are required to have a pseudorapidity difference between the
two \PH candidates $\abs{\Delta\eta} < 1.3$.

In the semi-resolved category, the ``pass'' region is defined by
the leading-\pt AK8 jet having a \Hbbt discriminant above $0.9$,
and the ``fail'' region below $0.9$.
The efficiency of the semi-resolved selection peaks at $\approx$3.5\%
around 1.2--1.4\TeV, depending on the signal, and rapidly falls
at higher \mx masses.
The requirements for the semi-resolved events are summarised in Table~\ref{tab:table2plus1}.

\begin{table}[ht]
	\centering
		\topcaption{Event selection criteria for the semi-resolved topology.}
		\label{tab:table2plus1}
		\begin{tabular}{ l c }
			\hline
			Variable                       & Selection                                       \\
			\hline
			Leading AK8 jet                & $\pt > 300 \GeV$ and $\abs{\eta} < 2.4$         \\
			AK4 jets                       & $\pt > 30 \GeV$ and $\abs{\eta} < 2.4$          \\
			$\abs{\Delta\eta}$             & $<$1.3                                         \\
			DeepJet                        & Medium working point                            \\
			Invariant mass of two AK4 jets & $90 < m_{\text{j}_{2} \text{j}_{3}} < 140\GeV$ \\
			$\mjjjs$                       & $>$750\GeV                                    \\
			\hline
		\end{tabular}
\end{table}

The main variable used in the search for an $\PH\PH$ resonance is the
``corrected $\PH\PH$ mass''.  For the fully-merged categories
it is defined as $\mjjs \equiv \mjj + (\mH - \mjone) + (\mH - \mjtwo)$,
where \mjj is the dijet invariant mass, \mjone and \mjtwo are the soft-drop masses of the
leading and subleading \PH candidate jets in the event, and $\mH =
	125\GeV$
is the nominal Higgs boson
mass.  In the semi-resolved analysis, this quantity is defined by
$\mjjjs \equiv m_{\text{Jjj}} + (\mH - \mjone) +
	(\mH - m_{\text{j}_{2} \text{j}_{3}})$, where $m_{\text{Jjj}}$ is the
invariant mass of the three jets comprising a semi-resolved $\PH\PH$ candidate.
The corrected $\PH\PH$ mass is used rather than the invariant mass of the two
reconstructed \PH candidates because effects due to fluctuations in jet
reconstruction or to missing \pt associated with a neutrino from a \PQb quark
decay are correlated between the \PH jet mass and the invariant mass of the
$\PH\PH$ system. Adjusting the \Hbb candidates to the nominal \PH mass
improves our estimate of the $\PH\PH$ invariant mass. Using the corrected $\PH\PH$ mass
leads to an 8--10\% improvement in the invariant jet mass
resolution~\cite{CMS:2017aza}. A requirement of $\mjjs > 750\GeV$ is applied
for selecting signal-like events because of trigger turn-on effects.

\section{Background model\label{sec:SigModelBkgEst}}

The background is predicted to be dominated by multijet and \ttjets production.
The contribution from processes like $\PW\PZ$, $\ttbar\PZ$ or $\PZ$+jets production
was found to be negligible.
The total background model is constructed as a sum of the individual
background contributions using a Poisson distribution for each bin of
the two-dimensional $(\mjone,\mjjs)$ distribution.
The \PH candidate jet mass is used as one of the discriminants because 
the signal and the \ttjets background exhibit shapes distinct from 
each other and from the multijet background.
To extract the signal, we compare the number of expected
events from both the background-only and signal-plus-background
hypotheses with the number of observed events in data using a likelihood ratio fit.
The number of extracted signal events can then be related to the production
cross section via
$N_\text{signal} = \sigma_X {\mathcal B}(\PX\to \PH\PH\to \bbbar\bbbar) \varepsilon L$,
where $\sigma_X$ is the production cross section of $\PX$ (a radion or a
bulk graviton), ${\mathcal B}(\PX\to \PH\PH\to \bbbar\bbbar)$ is the product of the
branching fractions of $\PX \to \PH\PH$ and the two \Hbb decays, $\varepsilon$ is the
product of the acceptance and the efficiency to reconstruct an $\PH\PH$
event, and $L$ is the integrated luminosity of the data set.

The multijet background estimation relies on a ``pass-to-fail ratio'',
a transfer function between the pass and fail regions defined in
Section~\ref{sec:EvtSel} and determined by the discriminant of the leading-\pt \PH candidate jet.
In this analysis, the pass-to-fail ratio is of the order of $10^{-2}$.
Conceptually, the pass-to-fail ratio is measured in the Higgs boson mass sidebands ($\mjone <
	100\GeV$ and $\mjone >140\GeV$) and interpolated into the signal region ($100<
	\mjone < 140\GeV$); however, both steps are done simultaneously with the
extraction of the signal yield and profiling over all nuisance parameters,
including those that govern the normalizations and shapes of the \ttjets component.

The \ttjets contributions to the signal categories are obtained from simulation, 
but their overall normalization and their shapes are allowed to be modified by 
nuisance parameters that are described in the next section. These nuisance parameters 
are constrained using 
two control event categories enriched in \ttjets, which are also a part of the joint
likelihood. Therefore, both the signal and all backgrounds are simultaneously
obtained from a one-step fit to the $(\mjone,\mjjs)$ planes in the ten pass and
fail regions.

The total numbers of expected events failing, $n_\text{F}$, and passing, $n_\text{P}$,
the \Hbbt requirement are given by
\begin{equation}
	n_\text{F} (i,\vec\theta) =
	n_\text{F}^\text{QCD}(i) + n_\text{F}^{\ttbar}(i,\vec\theta) + n_\text{F}^\text{\PX}(i,\vec\theta)
\end{equation}
and
\begin{equation}
	n_\text{P} (i,\vec\theta) =
	n_\text{P}^\text{QCD}(i) + n_\text{P}^{\ttbar}(i,\vec\theta) + n_\text{P}^\text{\PX}(i,\vec\theta)
\end{equation}
where $i$ is a bin in the 2D $(\mjone,\mjjs)$ plane, and
$\vec\theta$ is the set of all nuisance parameters that quantify the
systematic uncertainties, as described in Section~\ref{sec:Systematics}.
Each bin in the ``fail'' 2D distribution, $n_\text{F}^\text{QCD}(i)$,
is represented by an individual parameter in the fit that is required to be
positive but is otherwise unconstrained.

The predicted multijet yield in the ``pass'' 2D distribution,
$n_\text{P}^\text{QCD}(i)$, is obtained by
\begin{equation}
	n_\text{P}^\text{QCD}(i) = n_\text{F}^\text{QCD}(i) \, R_\text{P/F}(\mjone,\mjjs)
\end{equation}
where $R_\text{P/F}(\mjone,\mjjs)$ is the transfer function.

We define the transfer functions in data and in the QCD multijet simulation as
$R_\text{P/F}^\text{data}(\mjone,\mjjs)$ and $R_\text{P/F}^\text{sim}(\mjone,\mjjs)$,
respectively. The $R_\text{P/F}^\text{data}(\mjone,\mjjs)$ and
$R_\text{P/F}^\text{sim}(\mjone,\mjjs)$ both vary smoothly as a function of \mjone
and \mjjs because \PHH candidates in multijet processes arise from
random combinations of jets.  The
data-to-simulation ratio of these 2D functions,
\begin{equation}
	R_\text{ratio}(\mjone,\mjjs) \equiv \frac{ R_\text{P/F}^\text{data}(\mjone,\mjjs) }{ R_\text{P/F}^\text{sim}(\mjone,\mjjs)},
\end{equation}
is therefore also smooth and can be parameterised with an analytic function
of \mjone and \mjjs.

While $R_\text{P/F}^\text{data}(\mjone,\mjjs)$ could also be described by
analytic functions, features of this shape that are hard to model
analytically can be factored out by using the QCD simulation, and
the fit of the analytic function to data is only responsible for
describing the residual differences between data and simulation that
can be parameterised with fewer parameters than the shape of
$R_\text{P/F}^\text{data}(\mjone,\mjjs)$.
Thus the number of events in a given bin of the passing region is
obtained from
\begin{equation}
	n_\text{P}^\text{QCD}(i) = n_\text{F}^\text{QCD}(i) \, R_\text{P/F}^\text{sim}(\mjone,\mjjs) \, R_\text{ratio}(\mjone,\mjjs)
\end{equation}
where $R_\text{ratio}(\mjone,\mjjs)$ is a surface parameterised by the
product of two one-dimensional polynomials in the $(\mjone,\mjjs)$
plane with coefficients determined from the fit to data.
Second-order polynomials were chosen for $R_\text{ratio}(\mjone,\mjjs)$
parameterization, along both \mjone and \mjjs axes, based on a
Fisher test~\cite{Fisher}, where polynomial terms were added until the p-value
obtained in the test was larger than $0.05$.

To reduce the effect of statistical fluctuations on the calculation of
$R_\text{P/F}^\text{sim}(\mjone,\mjjs)$ in the QCD multijet simulation, the
pass and fail distributions are smoothed using an adaptive kernel
density estimate ~\cite{Cranmer:2000du} prior to calculating the ratio.

\section{Sources of systematic uncertainty}
\label{sec:Systematics}

The following sources of systematic uncertainty affect the expected signal and
background event yields. A complete list of systematic uncertainties and ranges
for the associated nuisance parameters is given in Table~\ref{tab:Syst}. These ranges
are used as input to the fit, and the minimization of the likelihood further
constrains some of them. None of these lead to a significant change in the
signal shape and, after the fit, their impact on the signal yield is
significantly smaller than the effect of limited statistics.

\begin{table}[ht!]
	\centering
	\topcaption{
		Summary of the impact of each source of systematic uncertainty on the signal and
		\ttbar background yields in the high purity signal region for a radion
		resonance at 1500\GeV. The impact of the same nuisance parameters in other 
		signal regions and for other resonance masses is similar.}
	\label{tab:Syst}
	{\renewcommand{\arraystretch}{1.2}
    \begin{center}
        \begin{tabular}{ l c c }
            \hline
            Source                            & Effect on signal (\%) & Effect on \ttbar background (\%) \\
            \hline
            Integrated luminosity             & 1.6                   & 1.6                               \\
            Pileup                            & 0.1                   & 0.2                               \\
            PDF and scales                    & 0.4                   & 1.2                               \\
            \ttbar cross section              & \NA                     & 5.0                               \\
            Trigger efficiency                & 4.0                   & 5.7                               \\
            Top quark $\pt$ reweighting      &\NA                     & 13.7                              \\
            DeepAK8 $\PH\to\bbbar$ efficiency & 18.3                  & 12.7                              \\
            DeepJet \PQb tagging efficiency   & 0.0                   & 0.0                               \\
            Jet energy scale                    & 1.5                   & 0.7                               \\
            Jet energy resolution             & 0.9                   & 0.9                               \\
            Jet mass scale                       & 1.2                   & 0.8                               \\
            Jet mass resolution               & 6.2                   & 6.7                               \\
            \hline
        \end{tabular}
    \end{center}
	}
\end{table}

The uncertainties in the modelling of the trigger response are particularly
important for $\mjjjs < 1100\GeV$, where the trigger efficiency drops below
99\%. 
The trigger efficiency in each category is measured in the data as a fraction
of events with at least one AK4 jet with $\pt > 260\GeV$ satisfying the offline
selection that passes the trigger selection criteria.
Simulated events are weighted by this efficiency as a function of the invariant mass of the two leading-\pt 
AK8 jets in the event, \mjjs.
The trigger efficiency in simulation is corrected by a scale factor,
which has an uncertainty between 1 and 15\%, attributable to the control
trigger inefficiency and the sample size used.

The impact of the jet energy scale and resolution
uncertainties~\cite{CMS-PAS-JME-16-003} on the signal yields is estimated to
be 1--3\%, depending on the signal mass. The jet mass scale and resolution are
measured using a sample of boosted $\PW \to {\qqbar'}$ jets in semileptonic
\ttbar events. The jet mass scale and resolution have a 2\% effect on the signal
yields because of a change in the mean of the \PH candidate jet mass distribution.

Scale factors are used to correct the signal event yields so that their \Hbbt and
DeepJet discriminant efficiencies are the same as for data. The \Hbbt and the
DeepJet discriminant scale factors are taken to be 100\% correlated. The
associated uncertainty in the scale factor is 2--9\%~\cite{Sirunyan:2017ezt},
depending on the \Hbbt working point and jet \pt, and is propagated to the total
uncertainty in the signal yield.

The impact of the uncertainties in the renormalization and factorization scale and
the parton distribution functions (PDF), the
latter derived using the {PDF4LHC} procedure~\cite{Butterworth:2015oua} and the
	{NNPDF3.0} PDF sets, is estimated to be 0.5\%. These uncertainties affect the
product of the signal acceptance and the selection efficiency. The factorization
scale and PDF uncertainties have negligible impact on the signal \mjjjs
distributions. Additional systematic uncertainties associated with pileup
modelling (1--2\%, based on a 4.6\% variation on the $\Pp\Pp$ total inelastic
cross section~\cite{CMSLumi13TeV}) and with the integrated luminosity
determination ~\cite{CMS:2021xjt,CMS-PAS-LUM-18-002,CMS-PAS-LUM-17-004}
(1.6\%, combining the measurements of the three years of data
taking),
are applied to the signal yield.

The systematic uncertainties applied to the signal are also applied to the \ttjets
background, as appropriate. The total uncertainty in the \ttjets cross section is
7\%.
The correction to the shape of the top quark $p_\mathrm{T}^\PQt$ distribution (described 
in Section~\ref{sec:Simulation}) has two parts, an additional normalization 
correction $\re^{\alpha}$ and a shape correction $\re^{-\beta p^\PQt_\mathrm{T}}$.
Each correction is assigned independent multiplicative uncertainties of 2
and 0.5 times the nominal weight.

An uncertainty in the ``bandwidth'' parameter of the
kernel density estimate,
which acts as a scale for the width of the adaptive kernels, was studied
by varying this parameter, and its impact is found to be negligible.

The main source of uncertainty in the multijet background estimate is the
statistical uncertainty in the fit of $R_\text{ratio}$. This uncertainty, amounting to 2--10\%, is fully correlated
between all \mjjjs bins. Additional statistical uncertainties in the background
shape and yield in the signal region result from the finite sizes of the
multijet samples in the fail region and are evaluated using the
Barlow--Beeston Lite method~\cite{BarlowBeeston,BBLite}. These uncertainties
are small compared to the uncertainty in the $R_\text{P/F}$ ratio, and are
uncorrelated from bin to bin.

\section{Results\label{sec:Results}}

\begin{figure}[!htbp]
	\centering
	\includegraphics[width=0.45\textwidth]{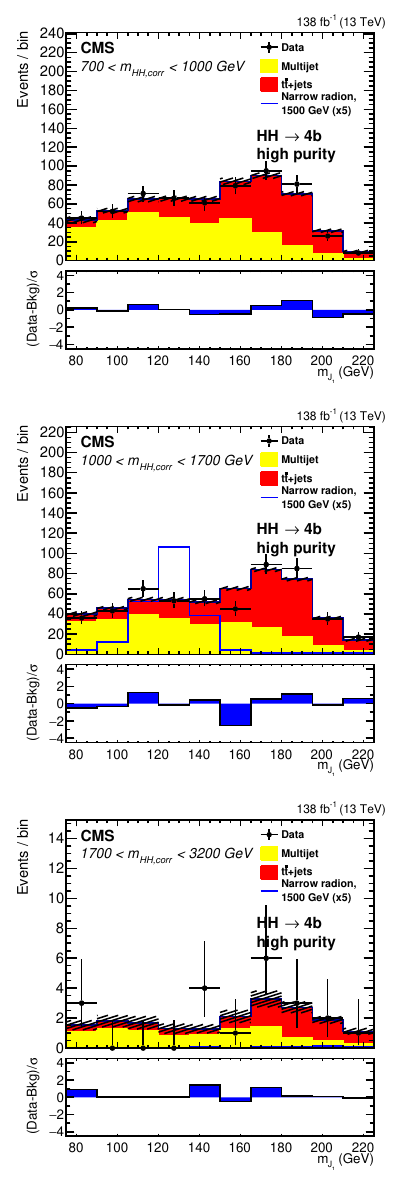}
	\includegraphics[width=0.45\textwidth]{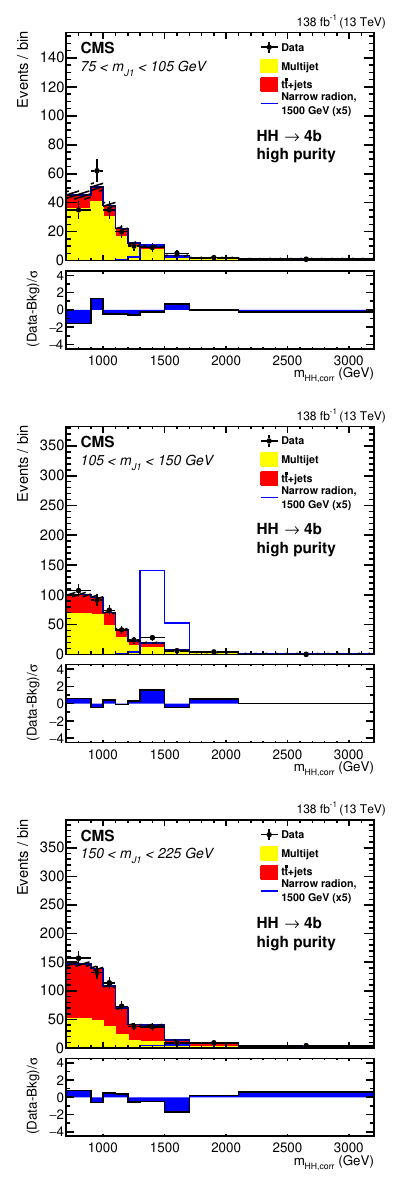}
	\caption{
		Slices of 2D distributions of observed events and the post-fit
		templates in the \TTcat signal region, projected onto the plane of leading jet mass
		$\mjone$ (left) and corrected $\PH\PH$ mass $\mjjs$ (right) axes,
		together with the signal (scaled up by a factor of five) 
		expected for a radion of mass 1.5\TeV. For this and following figures,
		the value of $\sigma$ in the lower panel is
		$\sigma = \sqrt{\sigma_\text{bkg}^2 + \sigma_\text{data}^2}$, where $\sigma_\text{bkg}$
		is the total post-fit uncertainty in the background and $\sigma_\text{data}$ is the statistical
		uncertainty associated with the number of data events in a particular bin.}
	\label{fig:HP}
\end{figure}

\begin{figure}[!htbp]
	\centering
	\includegraphics[width=0.45\textwidth]{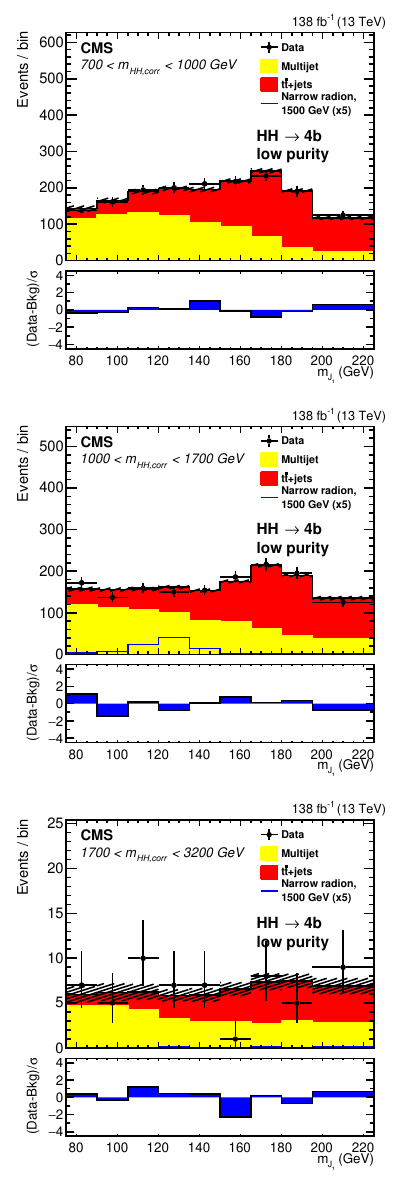}
	\includegraphics[width=0.45\textwidth]{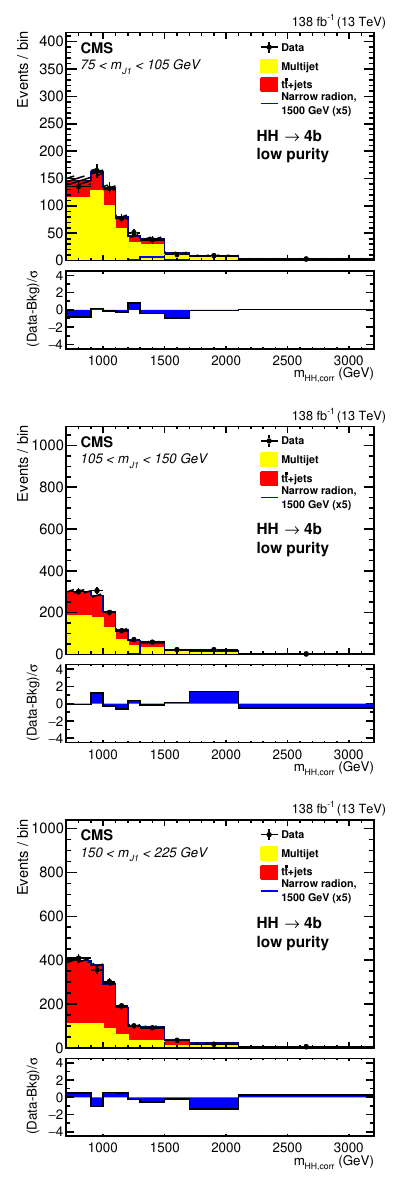}
	\caption{Slices of 2D distributions of observed events and the post-fit templates
		in the \LLcat signal region, projected onto the $\mjone$ (left) and $\mjjs$ (right) axes,
		together with the signal expected for a radion of mass 1.5\TeV. The signal is 
		scaled by the same factor as in Fig.~\ref{fig:HP}.}
	\label{fig:LP}
\end{figure}

\begin{figure}[!htbp]
	\centering
	\includegraphics[width=0.45\textwidth]{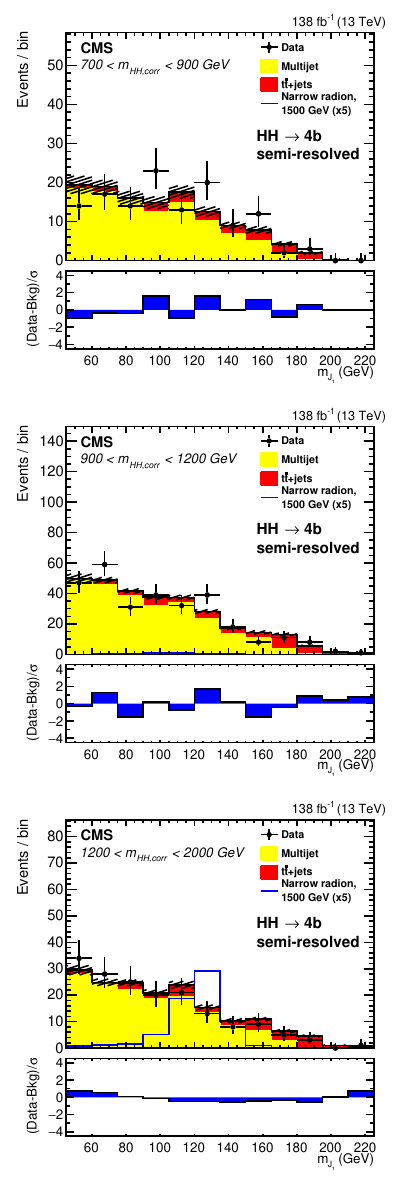}
	\includegraphics[width=0.45\textwidth]{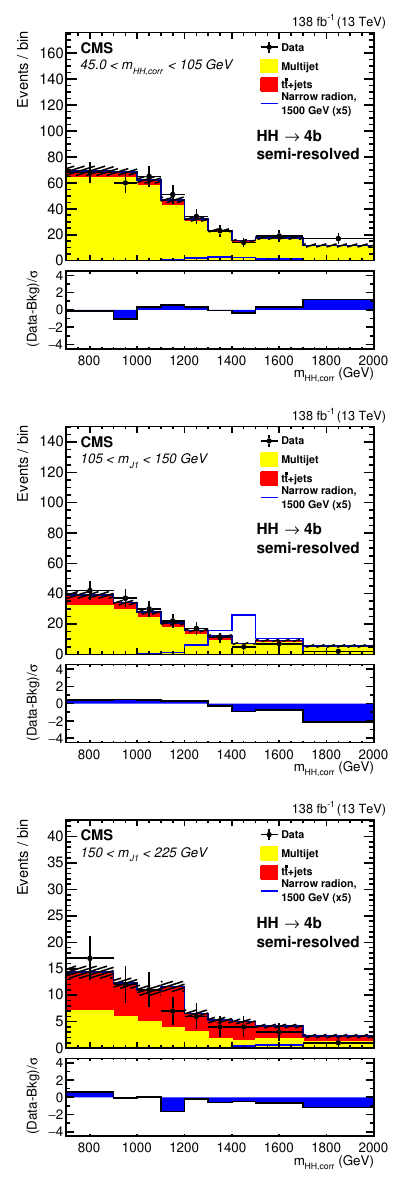}
	\caption{Slices of 2D distributions of observed events and the post-fit templates
		in the semi-resolved signal region, projected onto the $\mjone$ (left) and $\mjjs$
		together with the signal expected for a radion of mass 1.5\TeV. The signal is 
		scaled by the same factor as in Figs.~\ref{fig:HP} and~\ref{fig:LP}.}
	\label{fig:21}
\end{figure}

Results are obtained using a statistical combination of the semi-resolved and
fully-merged event categories. An $\PX \to \PH\PH$ signal is resonant in the 2D
space of the different signal event categories, as discussed in
Section~\ref{sec:SigModelBkgEst}. The likelihood is formed by combining
2D binned likelihoods of ten regions: \TTcat, \LLcat, and semi-resolved
signal categories, and \TTcat and \LLcat\ \ttbar control categories, where each category
provides both a pass and a fail region. The projections of the slices of the
post-fit 2D distributions in the three signal regions (\TTcat, \LLcat, and
semi-resolved) are shown in Figs.~\ref{fig:HP}--\ref{fig:21}.

The narrow radion signal corresponding to the resonance mass of 1500\GeV is
also shown. This resonance mass is chosen to illustrate the contribution of 
the semi-resolved category compared with the \TTcat and \LLcat categories; the 
signal is scaled by the same scale in all three figures.
The sensitivity is dominated by the \TTcat region over the whole
resonance mass domain. At lower resonance masses, the semi-resolved category
contributes significantly to the sensitivity of the search. The \LLcat category
contributes only at very high resonance masses, where the standard model
backgrounds are low.

The three signal regions are examined for an excess of
events above the predicted background, and the data are found to be
consistent with the expected background predictions.  We proceed
to set an upper limit on the number of possible signal events
in our data.

Upper limits at 95\% confidence level (CL) are set
on the product of the production cross section and the branching
fractions, $\sigma(\Pp\Pp\to\PX) \mathcal{B}(\PX\to \PH\PH\to \bbbar\bbbar)$.
They are obtained using the profile likelihood as a test
statistic~\cite{ATLASCMSComb}. The systematic uncertainties are
treated as nuisance parameters and are profiled in the minimization of
the negative of the logarithm of the profile likelihood ratio, and the
distributions of the likelihood ratio are calculated using the
asymptotic approximation~\cite{AsympCLs} of the procedure reported in
Refs.~\cite{CLS2,CLS1}.

\begin{figure}[thb!]
	\centering
	\includegraphics[width=0.49\textwidth]{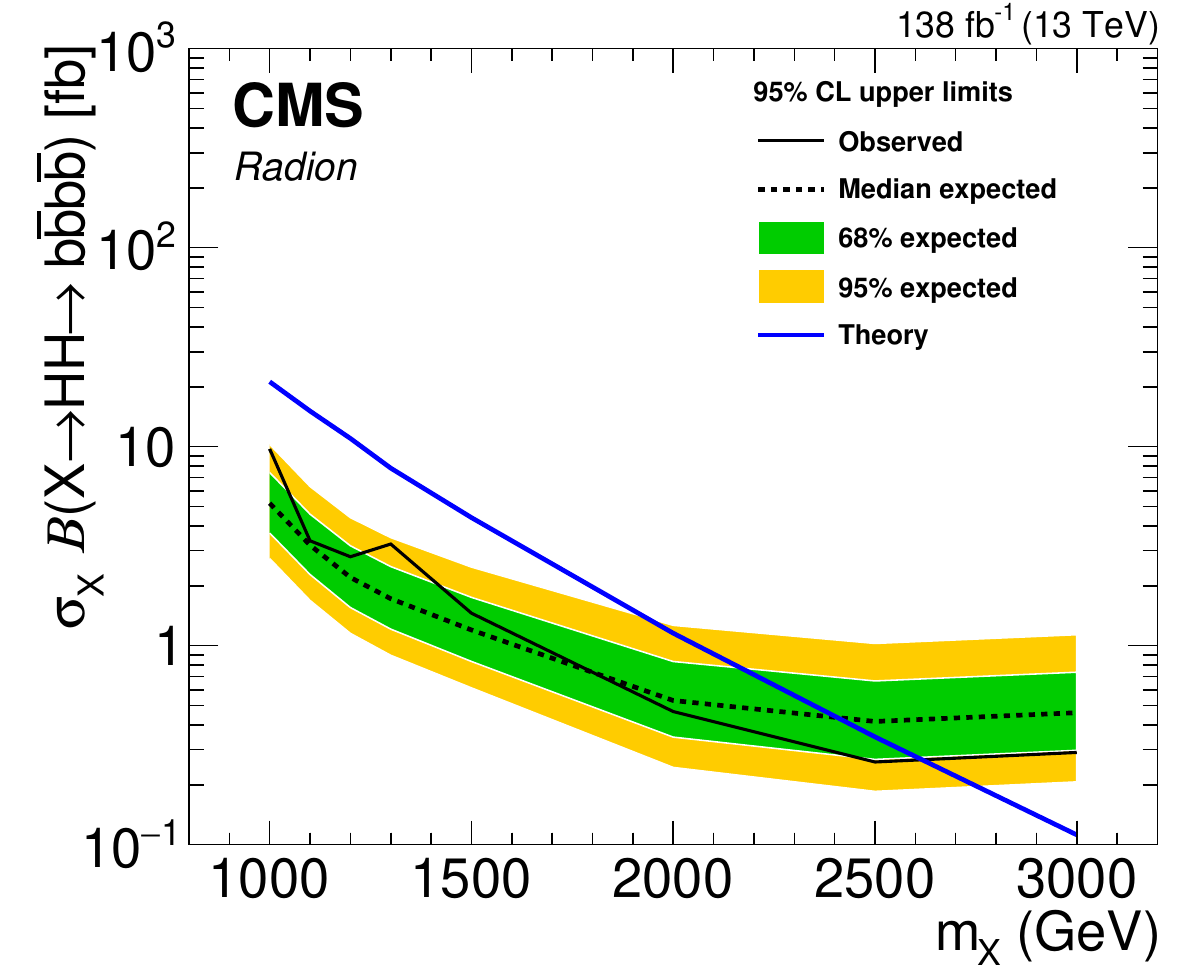}
	\includegraphics[width=0.49\textwidth]{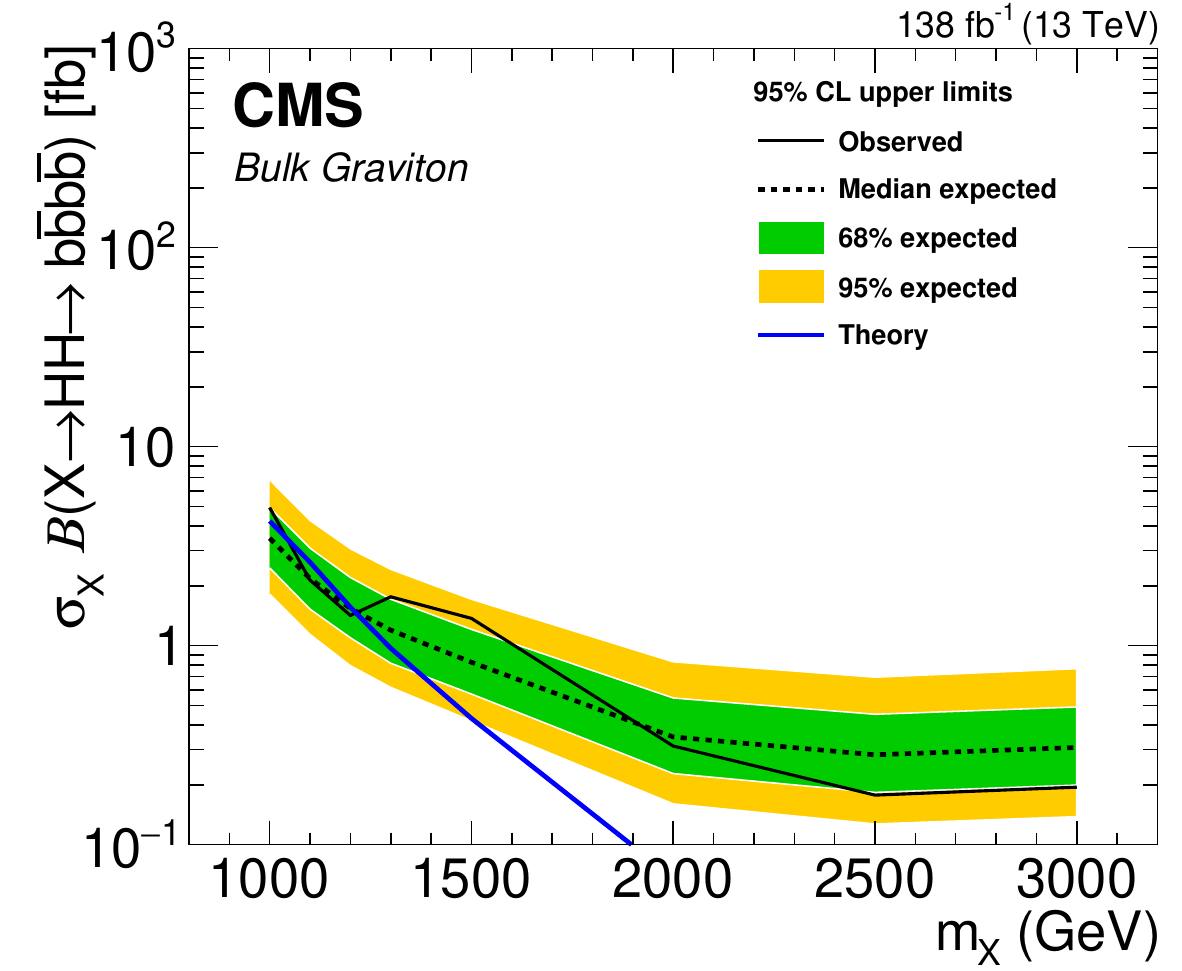}
	\caption{The observed (solid black line) and expected (dashed black line) upper limits at 95\% CL on
		$\sigma(\Pp\Pp\to\PX) \mathcal{B}(\PX\to \PH\PH\to \bbbar\bbbar)$
		for a narrow spin-0 radion
		(left, corresponding to $\LambdaR=3\TeV$) and
		a narrow width spin-2 bulk graviton
		(right, corresponding to $k/\Mplbar = 0.5$) models.
		The green (yellow) bands represent one (two) standard deviations from the expected limit.
		The predicted theoretical
		cross sections for the narrow radion and bulk graviton are also
		shown.}
	\label{fig:RadionLimitFULL}
	\label{fig:GravitonLimitFULL}
\end{figure}

\begin{figure}[thb!]
	\centering
	\includegraphics[width=0.49\textwidth]{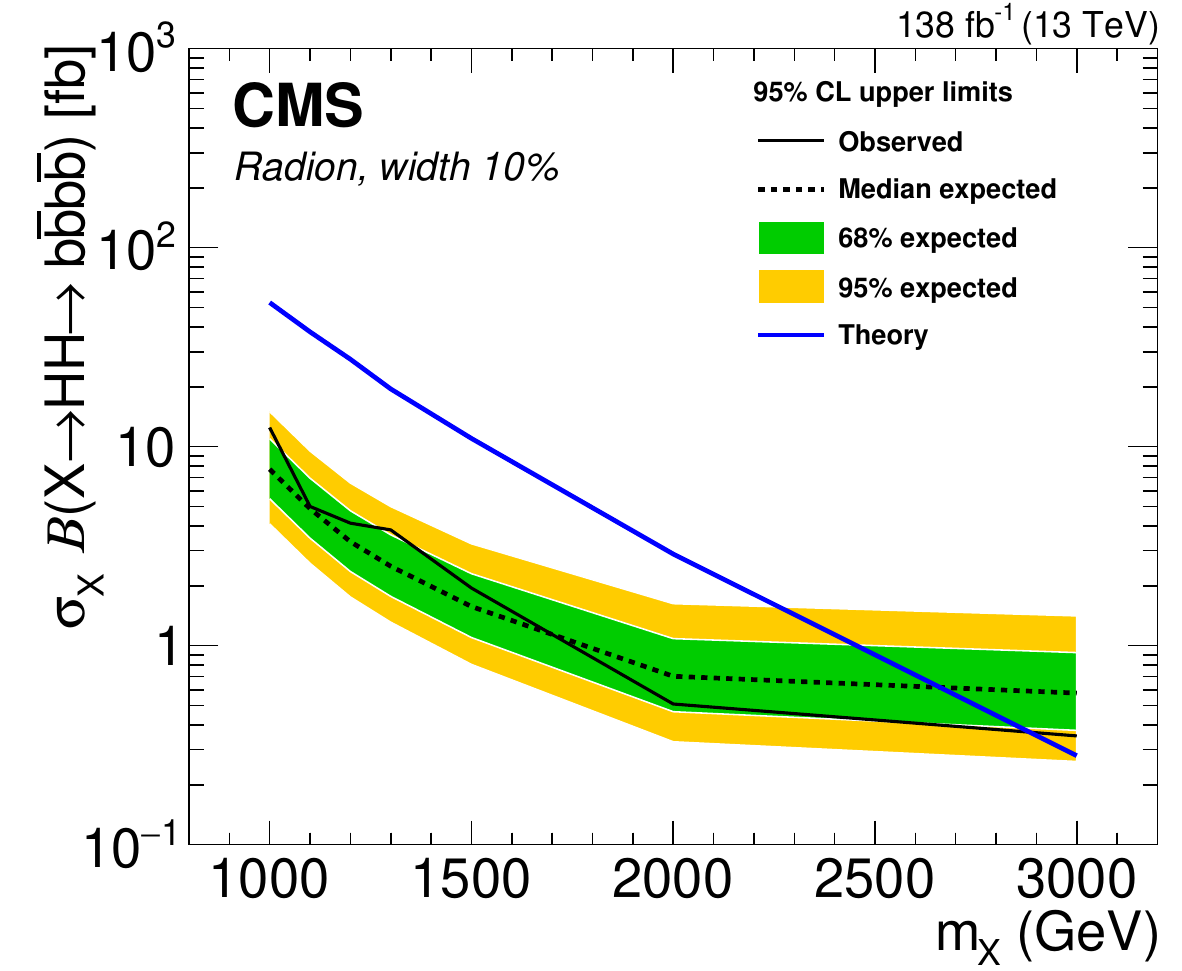}
	\includegraphics[width=0.49\textwidth]{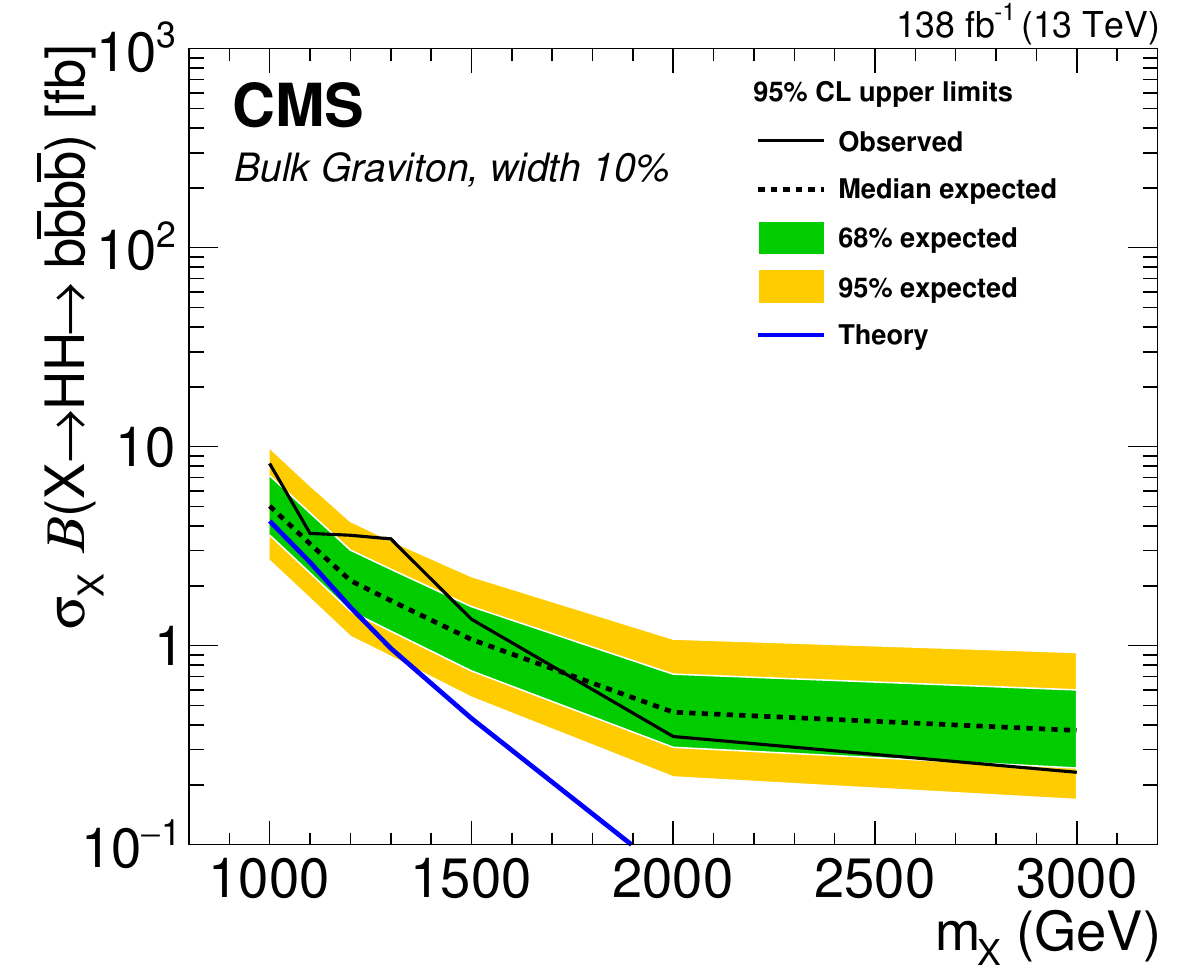}
	\caption{The observed (solid black line) and expected (dashed black line) upper limits at 95\% CL on
		$\sigma(\Pp\Pp\to\PX) \mathcal{B}(\PX\to \PH\PH\to \bbbar\bbbar)$
		for the 10\%-width spin-0 radion (left) and
		the 10\%-width spin-2 bulk graviton (right) models.
		The green (yellow) bands represent one (two) standard deviations from the expected limit.
		The predicted theoretical cross sections for the 10\%-width radion and bulk graviton are also shown.}
	\label{fig:RadionLimitFULLWide}
	\label{fig:GravitonLimitFULLWide}
\end{figure}

As shown in Fig.~\ref{fig:RadionLimitFULL} (left) a narrow radion with
mass between 1--2.6\TeV is excluded at 95\% CL
for the assumed value of the cutoff scale, $\LambdaR=3\TeV$.
A narrow bulk graviton for the assumed value of the ultraviolet cutoff scale,
$k/\Mplbar = 0.5$, is excluded at 95\% CL only for masses between 1--1.2\TeV,
as shown in Fig.~\ref{fig:GravitonLimitFULL} (right).
The deviations in the observed limits at graviton and radion masses of 1.3 and
1.5\TeV, respectively, correspond to a small upward fluctuation of data over
the background prediction at $\mjjs \approx$1.4\TeV, visible in
Fig.~\ref{fig:HP}, middle row.
The corresponding exclusion limits, assuming a signal with 10\% decay width,
are shown in Fig.~\ref{fig:RadionLimitFULLWide}.  The product of the efficiency 
and the acceptance for the 10\%-width signals is 3--5\% lower than for the 
narrow signals, and, consequently, the cross section exclusion limits are 
similar.

These limits result in the exclusion of the narrow-width graviton with \mx below 1.2\TeV.
Narrow and 10\%-width radion with masses below 2.6\TeV, and 2.9\TeV, respectively,
are also excluded.  This is a substantial improvement over the previous CMS radion
exclusion limit of $\approx$1.6\TeV\cite{CMS:2018vjd}.
The analysis presented in this paper complements a previous result from ATLAS
that achieved an almost identical sensitivity for $X \to \PH\PH \to 4\PQb$
for both spin-0 and spin-2 hypotheses~\cite{ATLAS:2022hwc} between 1.5--3\TeV, while employing a different
background estimation strategy and H jet identification.
Below 1.5\TeV the ATLAS analysis benefits from the combination with the fully
resolved $4\PQb$ channel.

\section{Summary\label{sec:Summary}}

A search has been presented for the pair production of standard model Higgs bosons
($\PH\PH$) from the decay of a spin-0 radion or a spin-2 bulk graviton as
predicted in warped extradimensional models, using data from proton-proton
collisions at a centre-of-mass energy of 13\TeV and corresponding to an
integrated luminosity of \intLumi.

The search is restricted to the
case where each Higgs boson decays to a bottom quark-antiquark pair.
It is conducted in the region of
phase space where at least one of the Higgs bosons has a large Lorentz
boost, so that the $\PH\to\bbbar$ decay products are collimated to
form a single \PH candidate jet.  The search combines events with one
$\PH$ candidate jet and two \PQb jets with events having two $\PH$ candidate jets,
thus adding sensitivity compared with previous
analyses~\cite{CMS:2017aza,CMS:2018vjd}.  The improvement 
comes from both an increase in integrated luminosity ($\approx$1.9)
and an improved \Hbbt ($\approx$2.5).

The results are interpreted in terms of upper limits on the
product of the production cross section for the respective resonance particles and the branching fraction
to $\PH\PH \to \bbbar\bbbar$, at 95\% confidence level. The upper limits range
from 9.74 to 0.29\unit{fb} for a narrow radion and from 4.94 to 0.19\unit{fb}
for a narrow bulk graviton, each having a mass of 1--3\TeV. Assuming a
width of 10\% for the radion and the graviton, the limits for the same masses are
in the range 12.48--0.35\unit{fb} and 8.23--0.23\unit{fb}, respectively.
As a result, the narrow-width graviton with \mx below 1.2\TeV, and
narrow and 10\%-width radion with masses below 2.6\TeV, and 2.9\TeV, respectively,
are excluded.

\begin{acknowledgments}

	We congratulate our colleagues in the CERN accelerator departments for the excellent performance of the LHC and thank the technical and administrative staffs at CERN and at other CMS institutes for their contributions to the success of the CMS effort. In addition, we gratefully acknowledge the computing centres and personnel of the Worldwide LHC Computing Grid and other centres for delivering so effectively the computing infrastructure essential to our analyses. Finally, we acknowledge the enduring support for the construction and operation of the LHC, the CMS detector, and the supporting computing infrastructure provided by the following funding agencies: SC (Armenia), BMBWF and FWF (Austria); FNRS and FWO (Belgium); CNPq, CAPES, FAPERJ, FAPERGS, and FAPESP (Brazil); MES and BNSF (Bulgaria); CERN; CAS, MoST, and NSFC (China); MINCIENCIAS (Colombia); MSES and CSF (Croatia); RIF (Cyprus); SENESCYT (Ecuador); ERC PRG, RVTT3 and MoER TK202 (Estonia); Academy of Finland, MEC, and HIP (Finland); CEA and CNRS/IN2P3 (France); SRNSF (Georgia); BMBF, DFG, and HGF (Germany); GSRI (Greece); NKFIH (Hungary); DAE and DST (India); IPM (Iran); SFI (Ireland); INFN (Italy); MSIP and NRF (Republic of Korea); MES (Latvia); LMTLT (Lithuania); MOE and UM (Malaysia); BUAP, CINVESTAV, CONACYT, LNS, SEP, and UASLP-FAI (Mexico); MOS (Montenegro); MBIE (New Zealand); PAEC (Pakistan); MES and NSC (Poland); FCT (Portugal); MESTD (Serbia); MCIN/AEI and PCTI (Spain); MOSTR (Sri Lanka); Swiss Funding Agencies (Switzerland); MST (Taipei); MHESI and NSTDA (Thailand); TUBITAK and TENMAK (Turkey); NASU (Ukraine); STFC (United Kingdom); DOE and NSF (USA).
	
	\hyphenation{Rachada-pisek} Individuals have received support from the Marie-Curie programme and the European Research Council and Horizon 2020 Grant, contract Nos.\ 675440, 724704, 752730, 758316, 765710, 824093, 101115353, 101002207, and COST Action CA16108 (European Union); the Leventis Foundation; the Alfred P.\ Sloan Foundation; the Alexander von Humboldt Foundation; the Science Committee, project no. 22rl-037 (Armenia); the Belgian Federal Science Policy Office; the Fonds pour la Formation \`a la Recherche dans l'Industrie et dans l'Agriculture (FRIA-Belgium); the F.R.S.-FNRS and FWO (Belgium) under the ``Excellence of Science -- EOS" -- be.h project n.\ 30820817; the Beijing Municipal Science \& Technology Commission, No. Z191100007219010 and Fundamental Research Funds for the Central Universities (China); the Ministry of Education, Youth and Sports (MEYS) of the Czech Republic; the Shota Rustaveli National Science Foundation, grant FR-22-985 (Georgia); the Deutsche Forschungsgemeinschaft (DFG), among others, under Germany's Excellence Strategy -- EXC 2121 ``Quantum Universe" -- 390833306, and under project number 400140256 - GRK2497; the Hellenic Foundation for Research and Innovation (HFRI), Project Number 2288 (Greece); the Hungarian Academy of Sciences, the New National Excellence Program - \'UNKP, the NKFIH research grants K 131991, K 133046, K 138136, K 143460, K 143477, K 146913, K 146914, K 147048, 2020-2.2.1-ED-2021-00181, and TKP2021-NKTA-64 (Hungary); the Council of Science and Industrial Research, India; ICSC -- National Research Centre for High Performance Computing, Big Data and Quantum Computing and FAIR -- Future Artificial Intelligence Research, funded by the NextGenerationEU program (Italy); the Latvian Council of Science; the Ministry of Education and Science, project no. 2022/WK/14, and the National Science Center, contracts Opus 2021/41/B/ST2/01369 and 2021/43/B/ST2/01552 (Poland); the Funda\c{c}\~ao para a Ci\^encia e a Tecnologia, grant CEECIND/01334/2018 (Portugal); the National Priorities Research Program by Qatar National Research Fund; MCIN/AEI/10.13039/501100011033, ERDF ``a way of making Europe", and the Programa Estatal de Fomento de la Investigaci{\'o}n Cient{\'i}fica y T{\'e}cnica de Excelencia Mar\'{\i}a de Maeztu, grant MDM-2017-0765 and Programa Severo Ochoa del Principado de Asturias (Spain); the Chulalongkorn Academic into Its 2nd Century Project Advancement Project, and the National Science, Research and Innovation Fund via the Program Management Unit for Human Resources \& Institutional Development, Research and Innovation, grant B05F650021 (Thailand); the Kavli Foundation; the Nvidia Corporation; the SuperMicro Corporation; the Welch Foundation, contract C-1845; and the Weston Havens Foundation (USA).	
\end{acknowledgments}

\bibliography{auto_generated}
\cleardoublepage \appendix\section{The CMS Collaboration \label{app:collab}}\begin{sloppypar}\hyphenpenalty=5000\widowpenalty=500\clubpenalty=5000
\cmsinstitute{Yerevan Physics Institute, Yerevan, Armenia}
{\tolerance=6000
A.~Tumasyan\cmsorcid{0009-0000-0684-6742}
\par}
\cmsinstitute{Institut f\"{u}r Hochenergiephysik, Vienna, Austria}
{\tolerance=6000
W.~Adam\cmsorcid{0000-0001-9099-4341}, J.W.~Andrejkovic, T.~Bergauer\cmsorcid{0000-0002-5786-0293}, S.~Chatterjee\cmsorcid{0000-0003-2660-0349}, K.~Damanakis\cmsorcid{0000-0001-5389-2872}, M.~Dragicevic\cmsorcid{0000-0003-1967-6783}, A.~Escalante~Del~Valle\cmsorcid{0000-0002-9702-6359}, R.~Fr\"{u}hwirth\cmsAuthorMark{1}\cmsorcid{0000-0002-0054-3369}, M.~Jeitler\cmsAuthorMark{1}\cmsorcid{0000-0002-5141-9560}, N.~Krammer\cmsorcid{0000-0002-0548-0985}, L.~Lechner\cmsorcid{0000-0002-3065-1141}, D.~Liko\cmsorcid{0000-0002-3380-473X}, I.~Mikulec\cmsorcid{0000-0003-0385-2746}, P.~Paulitsch, F.M.~Pitters, J.~Schieck\cmsAuthorMark{1}\cmsorcid{0000-0002-1058-8093}, R.~Sch\"{o}fbeck\cmsorcid{0000-0002-2332-8784}, D.~Schwarz\cmsorcid{0000-0002-3821-7331}, S.~Templ\cmsorcid{0000-0003-3137-5692}, W.~Waltenberger\cmsorcid{0000-0002-6215-7228}, C.-E.~Wulz\cmsAuthorMark{1}\cmsorcid{0000-0001-9226-5812}
% ...
\par}
\vskip\cmsinstskip
\dag:~Deceased\\
$^{1}$Also at TU Wien, Vienna, Austria\\
% ...
\end{sloppypar}
\end{document}